\documentclass{article}
\usepackage{icmc2025_paper_template}
\usepackage{times}
\usepackage{ifpdf}
\usepackage{soul}
\usepackage[english]{babel}
\usepackage{amsmath}
\usepackage{enumitem}
\usepackage{float}

\def\papertitle{Acoustic Wave Modeling Using 2D FDTD: Applications in Unreal Engine for Dynamic Sound Rendering}
\def\firstauthor{Bilkent Samsurya}
\def\secondauthor{Second Author}
\def\thirdauthor{Third Author}

\newif\ifpdf
\ifx\pdfoutput\relax
\else
   \ifcase\pdfoutput
      \pdffalse
   \else
      \pdftrue
  \fi
\fi

\ifpdf % compiling with pdflatex
  \usepackage[pdftex,
    pdftitle={\papertitle},
    pdfauthor={\firstauthor, \secondauthor, \thirdauthor},
    bookmarksnumbered, % use section numbers with bookmarks
    pdfstartview=XYZ % start with zoom=100% instead of full screen; 
   ]{hyperref}
  \usepackage[pdftex]{graphicx}
  \graphicspath{{./figures/}}
  \DeclareGraphicsExtensions{.pdf,.jpeg,.png}

  \usepackage[figure,table]{hypcap}

\else % compiling with latex
  \usepackage[dvips,
    bookmarksnumbered, % use section numbers with bookmarks
    pdfstartview=XYZ % start with zoom=100% instead of full screen
  ]{hyperref}  % hyperrefs are active in the pdf file after conversion

  \usepackage[dvips]{epsfig,graphicx}
  \graphicspath{{./figures/}}
  \DeclareGraphicsExtensions{.eps}

  \usepackage[figure,table]{hypcap}
\fi

\hypersetup{
    colorlinks,%
    citecolor=black,%
    filecolor=black,%
    linkcolor=black,%
    urlcolor=black
}

\title{\papertitle}

\oneauthor
   {\ Bilkent Samsurya} {Independent Researcher\\ %
     {\tt \href{mailto:bilkentcw@gmail.com}{bilkentcw@gmail.com}}}

\begin{document}
\capstartfalse
\maketitle
\capstarttrue
\begin{abstract}
Accurate sound propagation simulation is essential for delivering immersive experiences in virtual applications, yet industry methods for acoustic modeling often do not account for the full breadth of acoustic wave phenomena. This paper proposes a novel two-dimensional (2D) finite-difference time-domain (FDTD) framework that simulates sound propagation as a wave-based model in Unreal Engine, with an emphasis on capturing lower frequency wave phenomena, embedding occlusion, diffraction, reflection and interference in generated impulse responses. The process begins by discretizing the scene geometry into a 2D grid via a top-down projection, from which obstacle masks and boundary conditions are derived. A Python-based FD-TD solver injects a sine sweep at a source position, and virtual quadraphonic microphone arrays record pressure-field responses at pre-defined listener positions. De-convo-lution of the pressure responses yields multi-channel impulse responses that retain spatial directionality which are then integrated into Unreal Engine’s audio pipeline for dynamic playback. Benchmark tests confirm agreement with analytical expectations, and the paper outlines hybrid extensions aimed at commercial viability. 
\end{abstract}

\section{Introduction}\label{sec:introduction}
The current state of acoustic modeling sees the widespread use of geometric acoustics (GA) such as ray-tracing and beam-tracing to characterize the acoustic properties of virtual environments \cite{Savioja::01}. While these methods are generally considered efficient and accurate for modeling higher frequencies, GA models often require additional heuristics to capture lower frequency wave interactions~\cite{Savioja::01,Raghuvanshi::02}, prompting increased interest in exploring the use of more inclusive acoustic models.

Wave-based models offer a unified method of simulating sound propagation by directly solving the acoustic wave equation, which inherently incorporates diffraction, interference, and reflection throughout the audible frequency range ~\cite{Raghuvanshi::02,Hamilton::03}. One commonly employed wave-based technique is the finite-difference time-domain (FDTD) ~\cite{Hamilton::03,Ostashev::04}, which discretizes the time and spatial domain into a grid and iteratively solves for pressure wave propagation. Although this method can become computationally expensive for larger domains or longer simulation times, recent advances in parallel computing have made its commercial-scale application increasingly viable. 

In this paper, a proposal for designing a two-dimensional FDTD framework to model the acoustical response of a top-down outdoor environment in Unreal Engine is demonstrated. A sine sweep is used to acoustically excite the domain, allowing virtual microphone arrays at predefined listener positions to capture pressure field changes. The corresponding impulse responses are then extracted from the recorded pressure fields before being convolved with arbitrary sources during runtime. This approach facilitates seamless playback and a dynamic listening experience, offering more physically accurate diffraction and interference modeling than conventional geometric acoustics met-hods.

\section{The Finite-Difference Time-Domain (FDTD) Method}\label{sec:FDTD}
\subsection{Euler and Continuity Differential Equations}\label{subsec:body}

The conventional FDTD for acoustics is fundamentally ba- sed on the first-order Euler and continuity differential equations \cite{Jeong::05}, which in a lossless medium can be expressed as: 

\begin{equation}
\frac{\partial \mathbf{v}}{\partial t} = -\frac{1}{\rho_0} \nabla \cdot \ p
\label{eq:first order pressure derivative}
\end{equation}

\begin{equation}
\frac{\partial p}{\partial t} = -\rho_0 c^2 \nabla \cdot \mathbf{v}
\label{eq:first order ve;ocity derivative}
\end{equation}

where p is the sound pressure, v is the particle velocity vector, $\rho_0$ is the ambient density of the medium, c is the speed of sound in the medium. To solve these equations numerically, discretization is necessary. 
\subsection{Leapfrog Time Stepping and Staggered Spatial Grid}\
When adapted from electromagnetics to acoustics, Yee’s algorithm discretizes the wave equation in both space and time using a time stepping scheme and staggered spatial grid ~\cite{Zahari::06, Masoud::07}. For each time step, velocity is updated at half-integer time steps ($t + \Delta t /2$) and pressure fields at integer time steps ($t$). This temporal offset creates a 'leapfrogging' pattern between the two field updates ~\cite{Jeong::05,Zahari::06,Masoud::07}. In the spatial domain, a staggered spatial grid is employed such that for each grid cell, the pressure is stored in the cell center and velocity components are located at cell edges. Spatial derivatives are approximated using central difference methods ~\cite{Zahari::06,Masoud::07}. The discretized velocity and pressure update equations are given as: 

\begin{multline}
v_x^{n+\frac{1}{2}} \left(i+\frac{1}{2}, j \right) = v_x^{n-\frac{1}{2}} \left(i+\frac{1}{2}, j \right) - \\
\frac{\Delta t}{\rho \Delta x} \big(p^n(i+1, j) - p^n(i, j))
\end{multline}

\begin{multline}
v_y^{n+\frac{1}{2}} \left( i, j+\frac{1}{2} \right) = v_y^{n-\frac{1}{2}} \left( i, j+\frac{1}{2} \right) - \\
\frac{\Delta t}{\rho \Delta y} \left( p^n(i, j+1) - p^n(i, j) \right)
\end{multline}

\begin{multline}
p^{n+1}(i, j) = p^n(i, j) - \rho_0 c^2 \Delta t \cdot \\ 
\left[ \scalebox{0.8}{$\frac{V_x^{n+\frac{1}{2}} \left( i+\frac{1}{2}, j \right) - V_x^{n+\frac{1}{2}} \left( i-\frac{1}{2}, j \right)}{\Delta x}$} \right.
\left. + \scalebox{0.8}{$\frac{V_y^{n+\frac{1}{2}} \left( i, j+\frac{1}{2} \right) - V_y^{n+\frac{1}{2}} \left( i, j-\frac{1}{2} \right)}{\Delta y}$} \right]
\label{eq:discretepressure_update}
\end{multline}

where $n$ is the current time step, $x, y$ represent the 2D Cartesian coordinates,$v_x, v_y$ are the velocity components along the $x,y$ direction respectively, $\Delta 
t$ is size of the time step, $\Delta x, \Delta y$ are spatial step sizes along the $x,y$ direction respectively and $i,j$ denote the indices of the 2D grid along the $x$, $y$ axes respectively. Equations (3), (4), and (5) serve as the foundational update rules in the FDTD loop.

\begin{figure}[h]
\centering
\includegraphics[width=0.95\columnwidth]{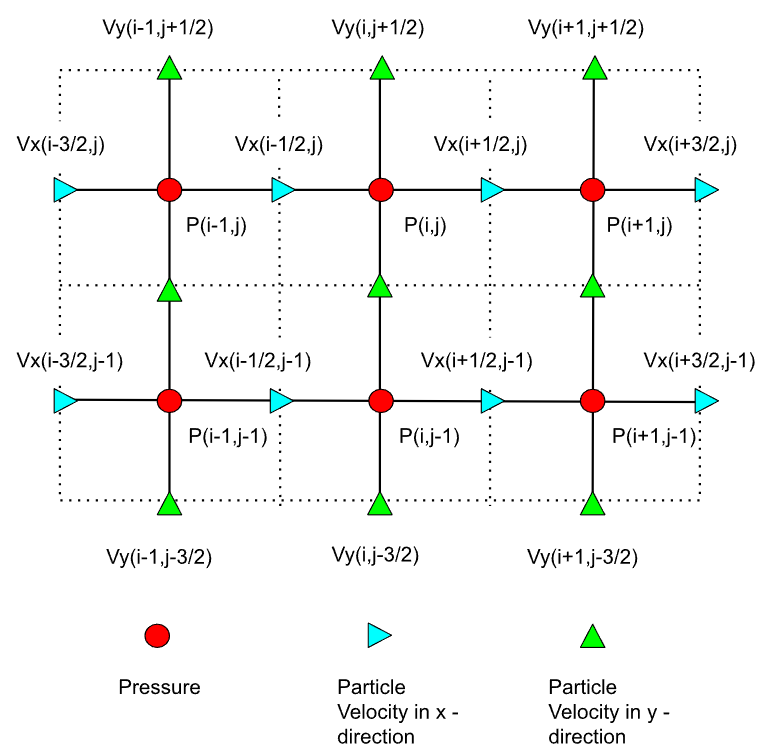}
\caption{Illustration of the staggered spatial grid, defining the spatial relationships between pressure and velocity fields. Pressure fields are located at the grid cell center and the velocity fields are positioned at the edge of each cell. \label{fig:example}}
\end{figure}

\section{Project Architecture}\label{sec:Project Architecture}

This acoustic simulation pipeline consists mainly of these four sequential stages:
\begin{enumerate}
    \setlength{\itemsep}{0pt}   
    \setlength{\parskip}{0pt}   
    \setlength{\parsep}{0pt}     
    \item Preprocessing in Unreal Engine
    \item Wave propagation simulation using FDTD in Python with sine sweep injection
    \item Impulse response extraction through deconvolution of pressure field recordings
    \item Real-time acoustic rendering in Unreal Engine by convolving pre-computed impulse responses with arbitrary source signals
\end{enumerate}

\section{Preprocessing in Unreal Engine}\label{subsec:body}
The test environment in Unreal Engine is conceptualized as an outdoor space modeled with a plane of dimensions $L_x = 32m$ meters by $L_y = 22m$ meters. Vowels of various configurations are generated on the surface of the plane to function as acoustical barriers. A single sound source is positioned at a fixed point, while multiple listener locations (L1 through L7) are distributed across the plane.

At each of these listener positions, a quadraphonic virtual microphone cluster is used to capture the acoustic responses in 4.0 surround sound. Conceptually, each ‘listener’ thus represents a tight four-microphone array, allowing for multi-channel spatial data collection. By exciting the environment with a sine sweep, the multi-channel recordings from each quadraphonic cluster are processed into impulse responses that capture both the spatial distribution and directional characteristics of the sound field defined by the given listener's orientation and location. 

\begin{figure}[h]
\centering
\includegraphics[width=0.9\columnwidth]{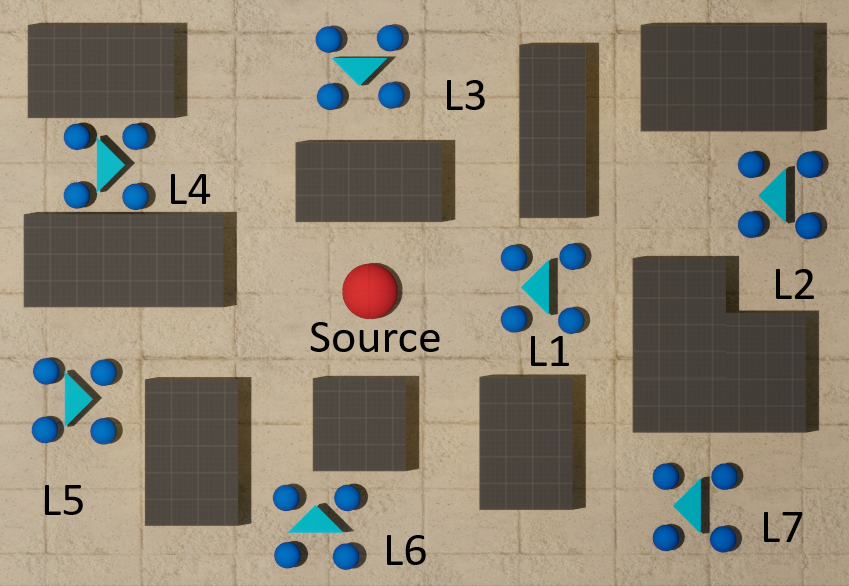}
\caption{The image represents the acoustic simulation domain where the sound source is denoted by a large red sphere positioned at the center. The listener locations, labeled L1 to L7, are marked with cyan triangles indicating their respective orientations. Each listener is surrounded by clusters of virtual microphones, represented by blue spheres, which capture spatialized acoustic data. The gray blocks scattered throughout the domain function as acoustic barriers\label{fig:example}}
\end{figure}

\subsection{Discretization}
The initial step in preprocessing involves discretizing the plane into a spatial grid. Each grid cell is defined by a spatial step size in both the \(x\) and \(y\) directions. In this implementation, the step sizes for the \(x\) and \(y\) directions are assumed to be equal, allowing the spatial step size to be expressed as a single variable, \(\Delta s\), where \(\Delta s = \Delta x = \Delta y\). It is shown in \cite{Zahari::06} that to reduce the likelihood of numerical dispersion, the spatial step should be at least a factor of 10 smaller than or equal to the shortest wavelength in the simulation. This implies that $\Delta s$ can be derived as: 

\begin{equation}
 \Delta s = 10 . \lambda_{{min}} = 10 .\frac{v_{max}}{f_{{max}}}
\label{eq:discretepressure_update}
\end{equation}

where $\lambda_{min}$ is the minimum wavelength, $v_{max}$ is the maximum speed of sound in all medium(s) of simulation and $f_{max}$ is the highest frequency to be captured in the simulation. The total number of spatial steps in each axis of the domain, $N_x$ and $N_y$, can then be obtained by dividing $L_x$ and $L_y$ by the previously derived spatial step size $\Delta s$.   

\subsection{Obstacle Mask Generation}
A classification process determines whether each grid cell contains an obstacle. The process begins by identifying acoustical obstacles through an \texttt{FGameplayTag} based framework, where each acoustical barrier receives a specific tag. A vertical line trace is then performed from the center of each grid cell in the positive z-direction. The grid cell is classified as 0 if the line trace intersects with a tagged obstacle, and 1 otherwise. This classification process, executed iteratively across all grid cells, generates an obstacle mask.

\begin{figure}[h]
\centering
\includegraphics[width=0.85\columnwidth]{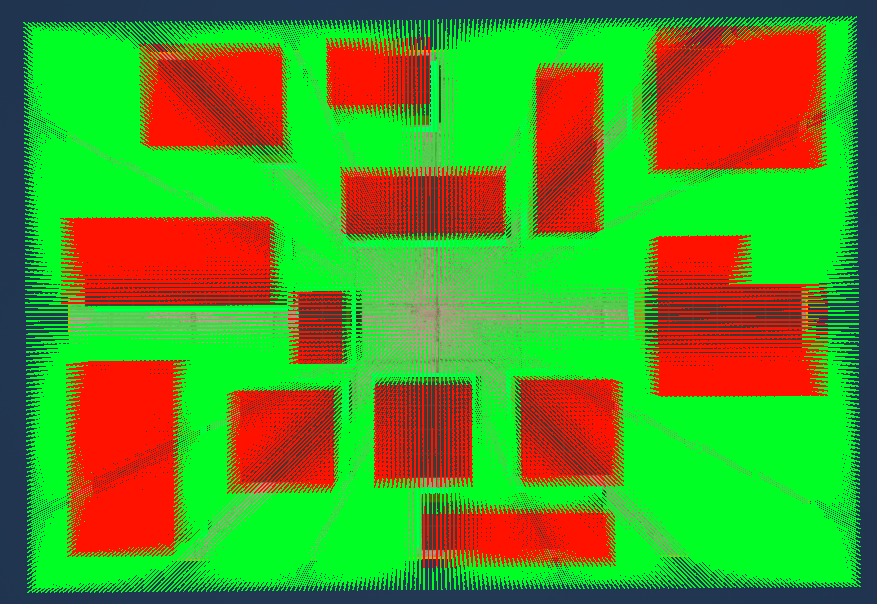}
\caption{Illustration of the discretization process. The (lighter) green lines represent no collisions by the line traces (free space) and the (darker) red lines indicate line trace intersections with acoustical barriers\label{fig:example}}
\end{figure}

\section{FDTD Implementation}
\subsection{Defining Time-Dependent Simulation Parameters}
For a given simulation time $T$, it follows that a stable time step size for a two dimensional FDTD simulation can be derived based on the Courant-Friedrichs-Lewy condition as ~\cite{Kowalczyk::08}:

\begin{equation}
 \Delta t \leq \frac{\Delta s}{c_{max} \sqrt{2}}
\label{eq:discretepressure_update}
\end{equation}

where $\Delta t$ is the size of the time step, $\Delta s$ is the spatial step size and $c_{max}$ is the maximum speed of sound accounting for all materials present in the simulation. Consequently, the total time steps  $N_t$ required for a simulation of time $T$ seconds can be determined by dividing $T$ by the time step size $\Delta t$. 

\subsection{Boundary Condition: PML}
One of the principal challenges in wave-based numerical simulations is properly replicating an unbounded domain. Traditional absorbing or non-reflective boundary conditio-ns can still produce unwanted reflections, undermining accuracy near the domain edges. To address this, a Perfectly Matched Layer (PML) formulation is adopted, originally introduced by Bérenger \cite{Berenger::09}, which provides minimal reflections while strongly attenuating outgoing waves.

PMLs are commonly defined by a thickness  of $N_{pml}$ cells that are applied at the perimeter of the simulation domain to manage boundary wave behavior. The gradual dissipation in this layer reduces reflections, mimicking free-field or open-air conditions.

In this implementation, the PML is characterized by a spatially varying damping coefficient, $\sigma_{x,y}(i,j)$, that transitions from a maximum value, $\sigma_{max}$, at the boundary to zero in the interior of the domain. Mathematically, the PML is constructed by separate one-dimensional ramps $\sigma_x(i)$ and $\sigma_y(j)$, each of width $N_{pml}$ cells on both ends of the domain in the x and y directions, respectively. The total damping in two dimensions is then obtained by summing these contributions:
\begin{equation}
\sigma_{x,y}(i,j) = \sigma_x + \sigma_y,
\end{equation}
with the damping terms for each axis defined as: 
\begin{multline}
\sigma_x(i) = \\
\begin{cases}
    \text{ramp from } \sigma_{\text{max}} \text{ to } 0, & 0 \leq i < N_{\text{pml}}, \\
    0, & N_{\text{pml}} \leq i \leq N_x - N_{\text{pml}}, \\
    \text{ramp from } 0 \text{ to } \sigma_{\text{max}}, & N_x - N_{\text{pml}} < i \leq N_x.
\end{cases}
\label{eq:piecewise}
\end{multline}
\begin{multline}
\sigma_y(j) = \\ 
\begin{cases}
    \text{ramp from } \sigma_{\text{max}} \text{ to } 0, & 0 \leq j < N_{\text{pml}}, \\
    0, & N_{\text{pml}} \leq j \leq N_y - N_{\text{pml}}, \\
    \text{ramp from } 0 \text{ to } \sigma_{\text{max}}, & N_y - N_{\text{pml}} < j \leq N_y.
\end{cases}
\label{eq:piecewise}
\end{multline}
where $\sigma_{x,y}(i,j)$ represents the two-dimension PML profile in both $x,y$ directions, $i,j$ represent the indices of the spatial grid on both $x,y$ directions respectively, $\sigma_{max}$ represent the maximum attenuation strength in the PML, $N_{x,y}$ represent the total number of spatial steps in the $x$ and $y$ directions of the domain and $N_{pml}$ represent the thickness (number of grid cells) of the PML. 

\subsection{Source Model: Transparent Source}
Impulse response recordings in acoustics commonly rely on an exponential (logarithmic) sine sweep (ESS) to excite the environment with equal energy per octave band ~\cite{Chan::10}. A key reason for employing an exponential sine sweep is its compatibility with Farina’s inverse filter (described in Section 7), allowing for clear separation of harmonic distortion and a clean extraction of the impulse response \cite{Farina::13}. In this FDTD scheme, the sweep is employed as a time domain forcing function and a transparent source, applied to the particle velocity fields at each time step. The transparent source (as opposed to a hard source) was chosen for the simulation as it adds energy incrementally to the field, allowing it to remain mostly unobtrusive to incoming waves. This approach avoids spurious reflections at the source and preserves the physical fidelity of the simulation \cite{Taflove::11}.

Spatially, the source is distributed over a small Gaussian footprint centered at a given location, rather than being confined to a single grid cell. Distributing the source over several cells smooths the injection into the grid, particularly at higher frequencies, and reduces abrupt spatial gradients that could otherwise lead to numerical artifacts \cite{Lin::12}. To ensure the source region spans at least a quarter-wavelength in all directions, or two grid cells, whichever is larger, the injection width is determined by:
\begin{equation}
 w_{x,y} = \max\left(2, \frac{\lambda_{min}}{2\Delta s}\right)
\label{eq:Source Width}
\end{equation}
where $w_{x,y}$ represents width of the source on both the $x$ and $y$ directions.

\subsection{FDTD Loop}
This section describes the time-stepping procedure in which the velocity and pressure fields are iteratively updated. During each time step, pressure values at the virtual microphone locations (see Figure 2) are recorded for subsequent impulse response extraction. The update loop consists of four main sequences:
\begin{enumerate}
    \setlength{\itemsep}{0pt}    % Space between items
    \setlength{\parskip}{0pt}    % Space between paragraphs
    \setlength{\parsep}{0pt}     % Space between paragraphs in items
    \item Velocity field updates
    \item Source injection
    \item Reflections from Obstacle
    \item Pressure field updates
\end{enumerate}

Referring back to Equation (3), the $x$ and $y$ components of the velocity field at each time step can be expressed in Python as:  

\begin{equation}
v_x \leftarrow v_x - \frac{\Delta t}{\rho \Delta x} \cdot \text{(p[1:, :] - p[:-1, :])}.
\label{eq:PressureDerivativeX}
\end{equation}

\begin{equation}
v_y \leftarrow v_y - \frac{\Delta t}{\rho \Delta y} \cdot \text{p[:,\,1:] - p[:,\,:\!-1])}.
\label{eq:PressureDerivativeY}
\end{equation}

To incorporate Perfectly Matched Layer (PML) boundary damping, 
the updated velocity fields \(\,v_x\)\, and \(\,v_y\)\, are multiplied by 
damping coefficients derived during the PML initialization step. 
The PML region is defined around the edges of the computational 
domain, and each cell within this region is assigned a specific damping factor 
to gradually attenuate outgoing waves and minimize artificial reflections 
from the boundaries. 

Next, a sine sweep source is introduced by injecting its time-varying 
pressure (or velocity) into the grid cells where the source is located. This 
procedure imposes a prescribed acoustic excitation that propagates through 
the domain according to the specified sweep function. 

Obstacle reflections are enforced by nullifying the 
velocity components \(\,v_x\)\, and \(\,v_y\)\, in the obstacle cells. 
The obstacle mask is generated from geometry data provided by Unreal Engine, 
where each obstacle cell is considered perfectly rigid. Setting the velocity 
to zero in these cells effectively simulates total reflection, thus preventing 
any acoustic waves from propagating through obstacle regions.

Finally, the pressure field is updated by incorporating the spatial derivatives of the velocity fields. These derivatives are computed as differences in the velocity components across grid cells:

\begin{equation}
\text{dp\_x} = \frac{v_{x}[1:, :] - v_{x}[:-1, :]}{\Delta x},
\label{eq:Velocity DerivativesX}
\end{equation}

\begin{equation}
\text{dp\_y} = \frac{v_{y}[:, 1:] - v_{y}[:, :-1]}{\Delta y}.
\label{eq:Velocity DerivativesY}
\end{equation}

Translated from Equation (5) into code form, the pressure field at the interior grid points is then updated using:

\begin{equation}
p[i,j] \leftarrow p[i,j] - \rho_0 c^2 \Delta t \left( \text{dp\_x} + \text{dp\_y} \right).
\label{eq:Pressure Update}
\end{equation}

This four-step sequence : velocity update, source injection, obstacle masking, and pressure update, is repeated for each time step, thereby propagating acoustic waves thro-ugh the simulation domain. At specified intervals, the pressure values at virtual microphone locations are stored, allowing for later analysis and the extraction of impulse responses.

\section{Simulation Outcomes and Interpretations}
\subsection{Pressure Maps}
The following normalized 2D acoustic traversal pressure maps represent the evolution of the pressure field in the Unreal Engine domain with dimensions $Lx = 32m$ by $Ly = 22m$ (refer to Figure 2) in progressive time steps for a period of 5 seconds. The domain was excited using a 2.5-second exponential sine signal with a frequency sweep from 20 Hz to 3 kHz placed at the center $(Lx/2, Ly/2)$ of the domain. The pressure maps represent the absolute normalized amplitude of the acoustic pressure in the domain. Red areas correspond to high-pressure zones or areas with high energy induction, while darker, blue region represents low-pressure zones or areas with minimal excitation. Acoustic obstacles are the dark blue voxelized areas in the map. They are taken to be rigid boundaries, exhibiting perfect reflection so there is no velocity and pressure changes in those zones.

\begin{figure}[htbp]
\centering
\includegraphics[width=1\columnwidth]{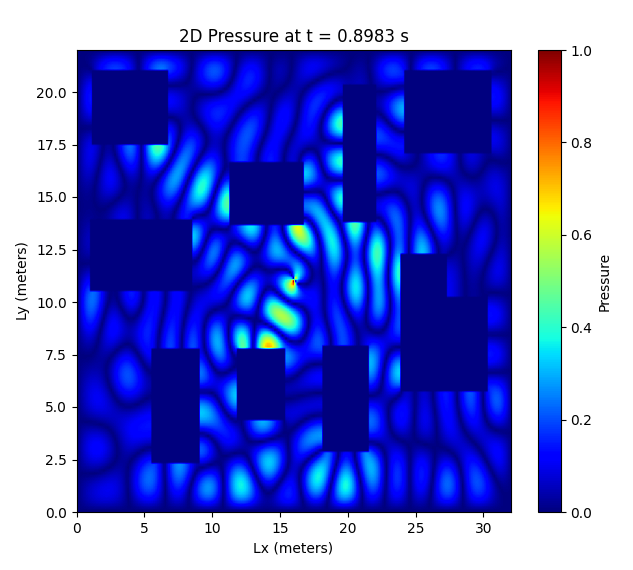}
\caption{Two‐dimensional pressure field at t = 0.8983s. At this early stage of the simulation, the initial wavefront has propagated outward from the source (indicated by the faint red glow at the center) and begun to interact with nearby obstacles. Reflections and partial shadowing of the wave can be seen where the acoustic field meets the rectangular blocks, and the color scale highlights regions of higher (yellow–red) and lower (blue) pressure amplitude.\label{fig:example}}
\end{figure}

\begin{figure}[H]
\centering
\includegraphics[width=1.0\columnwidth]{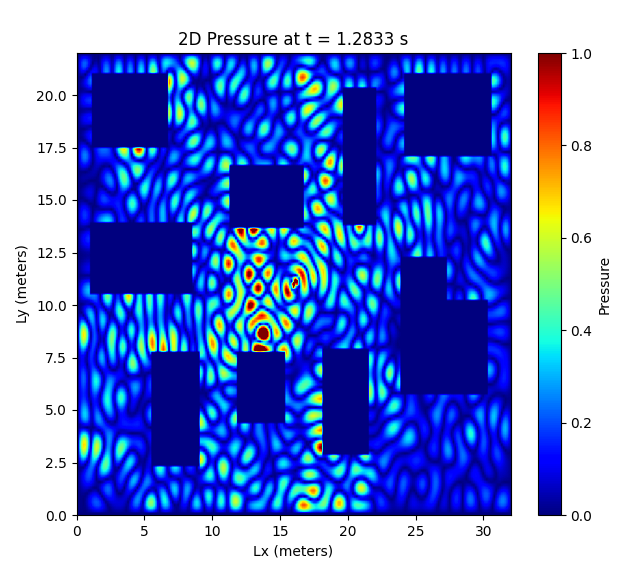}
\caption{Two‐dimensional pressure field at t = 1.2833s, showing more pronounced scattering and interference patterns around the obstacles. The ongoing sine sweep continues to excite the domain, producing higher amplitude regions (in red/yellow) and nodes (darker regions).\label{fig:example}}
\end{figure}

\begin{figure}[H]
\centering
\includegraphics[width=0.98\columnwidth]{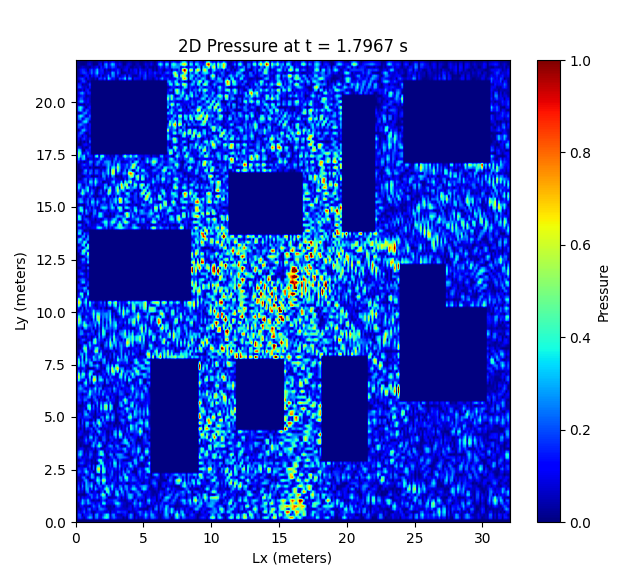}
\caption{Two‐dimensional pressure field at t = 1.7967s. As higher frequencies are swept, the shorter wavelengths resolve ever-smaller features, which manifests as a granular (speckle-like) interference pattern at later times due to the increasing contribution of high-frequency components.\label{fig:example}}
\end{figure}

\begin{figure}[htbp]
\centering
\includegraphics[width=0.99\columnwidth]{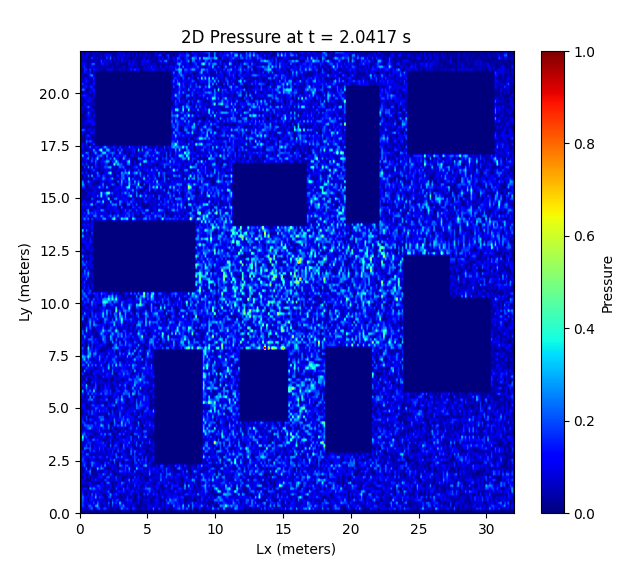}
\caption{Two‐dimensional pressure field at t = 2.0417s, near the end of the sine sweep. By this point, the field is dominated by interference of the various wavefronts reflecting from the perfectly reflecting obstacles. The near‐uniform blue background away from the scatterers indicates lower pressure amplitude regions as the sweep energy diminishes.\label{fig:example}}
\end{figure}

\subsection{Simulation Result Discussion}
\subsubsection{Evaluation of Pressure Maps}
In Figures 4–7, successive snapshots of the pressure field are illustrated within the computational domain as the sine sweep progresses and interacts with perfectly reflective obstacles (shown in dark blue). The rectangular barriers impose rigid boundary conditions, causing reflections and partial shadowing of the wavefronts. In Figure 4, the faint red glow near the center signifies the gradual accumulation of energy injected by the source; low‐frequency components dominate early on, giving rise to relatively smooth wavefronts that begin to reflect off the obstacles.

As the sine sweep transitions to higher frequencies in Figure 5, the wavefronts become more tightly spaced, leading to more complex interference patterns. Elevated pressure amplitudes (indicated by yellow–red regions) are particularly noticeable near the source, where overlapping wavefronts reinforce one another. By Figure 6, the sweep is nearing completion, and the intricate pressure distributions reveal the cumulative effects of reflection, diffraction, and interference around and between the rectangular blocks. The interplay of multiple frequencies is clearly visible in the interference fringes that form in the open regions of the domain.

At t = 2.0417s, Figure 7 shows that much of the acoustic energy has been absorbed by the perfectly matched layers (PMLs) at the boundaries and that new energy is being injected at a slower rate than the amount of energy the PMLs are absorbing. Only residual reflections dominate in the vicinity of the source and the obstacles, appearing as localized pockets of higher acoustic pressure. In this later snapshot, the wave has reverberated throughout the domain several times, creating a sparsely speckled field. Pressure amplitudes have become more evenly dispersed, reflecting a near-reverberant state where ongoing reflections continue but no distinct wavefronts visible.

\section{De-Convolution : Farina Inverse Filter}

\subsection{Resampling}
The recorded two-dimensional pressure field array essentially represent the domain’s acoustic response to the sine sweep. At each of the predefined listener positions (\( L_1 \) – \( L_7 \)), contains a cluster of four equally spaced virtual omnidirectional microphones. The pressure field changes at each of the microphone locations are captured for subsequent deconvolution, thereby extracting the impulse responses at their respective locations. With \( N_t = 214,285 \) total time steps, the effective sampling rate is \( f_s = 42,857 \) Hz. Since the highest frequency in the sine sweep is \( f_{\max} = 3,000 \) Hz, it follows that $f_s > 2 f_{\max}$ satisfying the Nyquist criterion. To facilitate futher audio processing, the signals can then be safely resampled to the standard audio rate of \( 44.1 \) kHz.

\subsection{Farina’s Inverse Filter for Exponential Sine Sweep (ESS) De-Convolution} 
When a sine sweep is injected into a system with an impulse response $h(t)$, the recorded signal $r(t)$ can be given as the convolution of the sine sweep and impulse response: 
\begin{equation}
 r(t) = s(t) \ast h(t)
\label{eq:Source Width}
\end{equation}
where $r(t)$ is the recorded signal, $s(t)$ is the sine sweep and $h(t)$ represents the impulse response. To solve for the impulse response, Farina gives us a three-step procedure to formulate the inverse filter \cite{Farina::13}: 
\begin{enumerate} 
    \setlength{\itemsep}{0pt}    % Space between items
    \setlength{\parskip}{0pt}    % Space between paragraphs
    \setlength{\parsep}{0pt} 
    \item  Time-reverse the original sweep signal,
    \item  Apply an exponential amplitude correction term to the time reversed sweep,
    \item  Convolve the result with the recorded signal. 
\end{enumerate}
Mathematically, the derivation of the system's impulse response from the recorded signal can be represented as: 
\begin{equation}
 h(t) = r(t)\ast s(T-t)e^{-\alpha (T - t)}
\label{eq:Source Width}
\end{equation}
given that: 
\begin{equation}
\small\alpha = \frac{\ln\left(\frac{f_1}{f_0}\right)}{T}
\label{eq:Source Width}
\end{equation}
where $s(T-t)$ represents the time-reverse sine sweep, $T$ is the duration of the sweep, $f_1$ and $f_0$ are the end and start of the sweep's frequency values respectively. Figure 9 below shows the rendered four channel impulse response file after de-convolution: 

\begin{figure}[h]
\centering
\includegraphics[width=0.4\columnwidth]{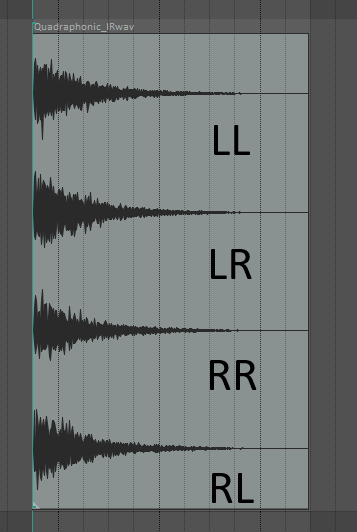}
\caption{The wave file of the quadraphonic impulse response (IR) corresponding to listening position L2. Each of the four channels is labeled to reflect the channel configuration of a true stereo IR format. The true stereo IR concept will be explained in the next section.\label{fig:example}}
\end{figure}

\section{Implementation in Unreal Engine}

\subsection{True Stereo Impulse Responses} 
With each of the listener positions representing a four microphone array, four separate mono impulse responses were extracted for a single reference point. The motivation behind capturing four tracks is to approximate a form of true stereo impulse response in a single pass. Conventionally, a true stereo impulse response is a quadraphonic (four-channel) recording that preserves the inter-aural time differences and directional cues of the reflections, thereby improving the sense of localization and depth in the stereo field \cite{Steinberg::14}. 

A typical true stereo capture can be achieved by performing two separate recordings to obtain two stereo files \cite{SIR::15}:
\begin{enumerate} 
    \setlength{\itemsep}{0pt}    
    \setlength{\parskip}{0pt}    
    \setlength{\parsep}{0pt} 
    \item  First pass: The source is placed closer to one microphone (e.g., the left), and a test signal (such as a sine sweep) is played and recorded.
    \item  Second pass: The source is then placed closer to the other microphone (the right), and the same test signal is recorded once more.
\end{enumerate}

From these two passes, a four-channel impulse response can be derived:
\[
\begin{aligned}
    \text{Left}_{\text{Source}} &\to \text{Left}_{\text{Channel}} ( L\to L)\\
    \text{Left}_{\text{Source}} &\to \text{Right}_{\text{Channel}} ( L\to R) \\
    \text{Right}_{\text{Source}} &\to \text{Left}_{\text{Channel}} ( R\to L) \\
    \text{Right}_{\text{Source}} &\to \text{Right}_{\text{Channel}} ( R\to R)
\end{aligned}
\]
These four channels capture how the sound propagates differently when the source is near one microphone versus the other, preserving directional information and enhancing the spatial characteristics in subsequent playback.

In the present work, instead of running two separate FDTD simulations (one for each source position), a single simulation is performed while placing four microphones at fixed positions around the listening point. By doing so, the equivalent of these two passes is effectively captured in one go, because the FDTD grid simultaneously computes sound propagation to all microphones in the model. Each microphone's position and orientation are determined by its cluster's spatial relationship to the listener. This arrangement categorizes the microphones into four groups: front-left, front-right, rear-left, and rear-right. In the context of a "true stereo" structure:
\begin{enumerate}
    \setlength{\itemsep}{0pt}    
    \setlength{\parskip}{0pt}    
    \setlength{\parsep}{0pt}     
    \item Front-left corresponds to L → L
    \item Rear-right corresponds to L → R
    \item Front-right corresponds to R → R
    \item Rear-left corresponds to R → L
\end{enumerate}
The channels in the quadraphonic WAV file are structured in this same sequence, ensuring proper spatial alignment in the recording. Although it is not an exact reproduction of the standard two-pass technique, it preserves many of the spatial benefits, offering improved localization and depth compared to a conventional single-channel or single-stereo impulse response.

\subsection{Head Orientation and Energy Distribution Across IR Channels} 
As the listener’s orientation changes in Unreal Engine, the 3D panner of a given sound source sends more or less of the source signal into the left or right inputs of the true‐stereo convolution reverb. This causes the relative energy in each of the four IR channels (L→L, L→R, R→L, R→R) to shift accordingly. For example, if the listener’s right ear is facing the source, the panner emphasizes the right input, making the front‐right (R→R) and rear‐left (R→L) channels more prominent in the resulting reverb. Conversely, if the listener faces the source directly, the panner balances the source signal between both inputs, so all four IR channels contribute more evenly. This of course assummes the source is 3D and pannable. Figure 10 establishes the relationship between head orientation and energy distribution in the four channel true stereo impulse response for a given 3D source.

\begin{figure}[h]
\centering
\includegraphics[width=0.8\columnwidth]{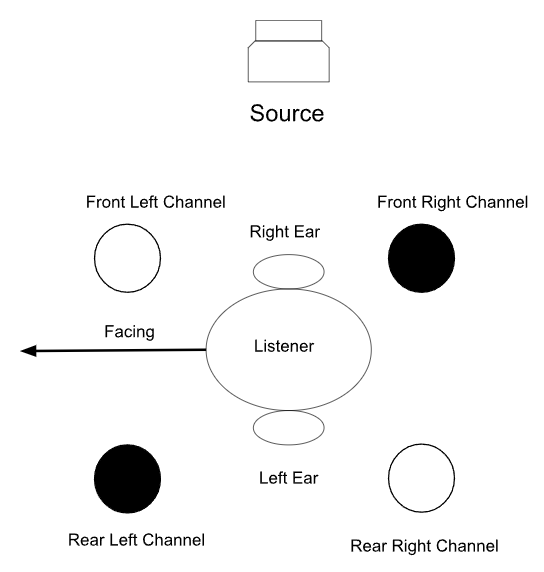}
\caption{Illustration of the relationship between head orientation and energy distribution in the quadraphonic, true stereo impulse response (IR) in Unreal Engine. When the listener’s right ear is facing the source, most of the IR energy contribution comes from the front-right channel and rear-left channel indicated by the shaded circles. \label{fig:example}}
\end{figure}.

Each of the four-channel impulse responses (IRs) represents the acoustic characteristics of a specific region within the environment for a corresponding source location. During playback, these IR clusters are categorized as either active or inactive, depending on their relevance to the listener's position. In Unreal Engine (UE), these impulse responses are implemented as an effect preset \texttt{SubmixEff-\newline ectConvolutionReverbPreset}, which is assigned to the \texttt{SubmixSends} array in the source configuration. 

At any given time, two active four-channel IRs are convolved with the source, meaning the Submix Sends array contains two active elements. The send amounts for these active IRs are dynamically weighted based on the listener's position relative to them. The contribution of each cluster is determined by the listener’s distance to both active clusters—moving closer to Cluster A increases its contribution while reducing Cluster B’s contribution.

If the send amount of one cluster falls below a predefined threshold, the system identifies the next closest cluster and replaces the least relevant one. This allows the Submix Sends array to function as a circular buffer, ensuring a smooth and continuous transition between impulse responses as the listener moves through the environment.

\section{Accuracy Analysis and Benchmarks}
A comparison of the two-dimensional FDTD response with an analytical solution constructed using the free space Gre-en's function\footnote{Green's function enables the calculation of acoustic fields at any point in the domain by integrating over source distributions. It serves as a fundamental reference for validating analytical solutions to wave equations in complex acoustic environments.}was performed to verify the FDTD precision. For each test, a Ricker wavelet of center‐frequency $f_0$ is injected at the source grid cell; the analytical impulse response at the receiver is obtained by convolving the Green’s function with the same Ricker pulse. Both the numerical and analytical traces are sampled in equal time step and at the same source-listener locations to isolate dispersion‐driven errors as a function of frequency. The normalized root‐mean‐square error and the peak‐arri-val‐time difference was computed at each center-frequency. This frequency‐dependent error characterization demonstrates that the grid‐spacing criterion maintains amplitude errors below 4$\%$ and phase error below $1.5  ms$ across the operating frequency range
\begin{figure}[h]
\centering
\includegraphics[width=1.0\columnwidth]{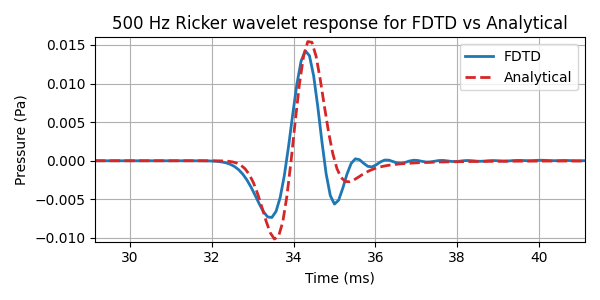}
\caption{ \label{fig:example} Simulated (FDTD) vs Analytical time-domain pressure responses to a 500Hz Ricker pulse input signal}
\end{figure}.

\begin{table}[h]
  \centering
  \begin{tabular}{|l|l|l|l|}
    \hline
    $f_0$ Hz
  & NRMSE (\%)
  & Arrival Time (ms) \\
    \hline
    250 & 2.3 & 0.2\\
    500 & 3.6 & 0.5\\
    1 K & 3.7 & 1.1\\
    3 K & 3.9 & 1.2\\
    \hline
  \end{tabular}
  \caption{Frequency-dependent validation results. For each Ricker wavelet center-frequency \(f_0\) (250 Hz, 500 Hz, 1 kHz, 3 kHz), the table reports the normalized root-mean-square error (NRMSE, \%) and the peak-arrival-time difference \(\Delta t\) (ms) against the analytical Green’s-function solution}
  \label{tab:example}
\end{table}
\begin{figure}[h]
\centering
\includegraphics[width=1\columnwidth]{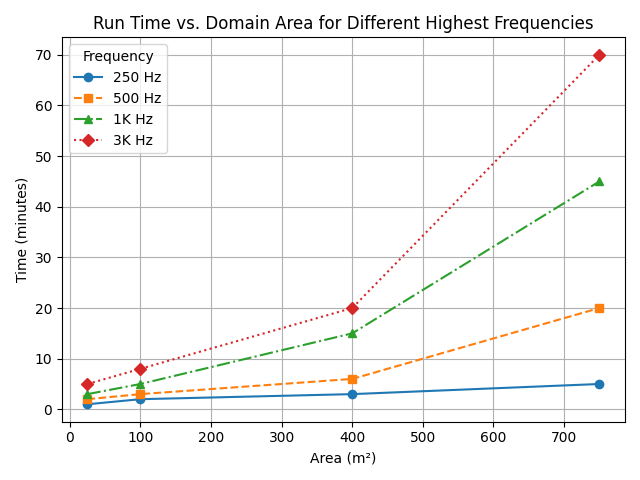}
\caption{ \label{fig:example} 2D FDTD simulation run time (minutes) plotted against computational domain area (m²), for four different highest frequencies sampled (250 Hz, 500 Hz, 1 kHz, 2 kHz).}
\end{figure}

\section{Constraints and Improvements}
A constraint for this method is that to capture higher frequency ranges, the complexity and resulting processing time becomes significant even with parallel computing. The amount of spatial resolution needed and thus number of spatial grid can also incur longer simulation time. A remedy and improvement to the current workflow might be to incorporate ray or beam tracing for higher frequency response capture, hybridizing the two workflows and enabling the distribution of computational complexity over two models. 

\section{Conclusion}
The proposed workflow demonstrates a working model of a wave-based FDTD acoustic solver in Unreal Engine, yiel-ding physically accurate low-frequency sound modeling. Advances in parallel computing have increased FDTD’s efficiency but to achieve full-spectrum sound rendering suitable for production use, a hybrid scheme is recommended: FDTD handles lower frequency wave phenomena, while geometric methods captures high frequency content. Such a division preserves FDTD’s fidelity where it is most need-ed and leverages the efficiency of geometric acoustics for short wavelength behavior, significantly reducing overall computation time. This complementary pairing provides a scalable, high-fidelity acoustic rendering solution for gam-es, VR/AR, and architectural simulations.

%%%%%%%%%%%%%%%%%%%%%%%%%%%%%%%%%%%%%%%%%%%%%%%%%%%%%%%%%%%%%%%%%%%%%%%%%%%%%
%bibliography here
\bibliography{icmc2025_paper_template}

\end{document}